%% file: paper.tex
\documentclass[3p,times,preprint]{elsarticle}

\include{commonm}
\usepackage{hyperref}

\journal{ }

\begin{document}

\begin{frontmatter}

\title{Comparison of methods for curvature estimation from volume fractions}

\author[cornell]{Austin Han\corref{mycorrespondingauthor}}
\ead{ah2262@cornell.edu}
\author[cornell,ovgu]{Fabien Evrard}
\author[cornell]{Olivier Desjardins}

\cortext[mycorrespondingauthor]{Corresponding author}

\address[cornell]{Sibley School of Mechanical and Aerospace Engineering, Cornell University, Ithaca, NY 14853, United States of America}
\address[ovgu]{Lehrstuhl f\"{u}r Mechanische Verfahrenstechnik, Otto-von-Guericke-Universit\"{a}t Magdeburg, Universit\"{a}tsplatz 2, 39106 Magdeburg, Germany}

\begin{abstract}
This paper evaluates and compares the accuracy and robustness of curvature estimation methods for three-dimensional interfaces represented implicitly by discrete volume fractions on a Cartesian mesh. The height function (HF) method is compared to three paraboloid fitting methods: fitting to the piecewise linear interface reconstruction centroids (PC), fitting to the piecewise linear interface reconstruction volumetrically (PV), and volumetrically fitting (VF) the paraboloid directly to the volume fraction field. 
The numerical studies presented in this work find that while the curvature error from the VF method converges with second-order accuracy as with the HF method for static interfaces represented by exact volume fractions, the PV method best balances low curvature errors with low computational cost for dynamic interfaces when the interface reconstruction and advection are coupled to a two-phase Navier-Stokes solver.
\end{abstract}


\end{frontmatter}


\section{Introduction}
The calculation of surface tension forces in interfacial flows requires the accurate estimation of the phase interface mean curvature to avoid spurious flow near the interface induced by curvature estimation errors. Curvature calculation presents a challenge within the context of a volume-of-fluid (VOF) scheme \cite{DeBar1974FundamentalsCode,Noh1976SLICCalculation,Nichols1980SOLA-VOF:Boundaries} because of the implicit and discontinuous nature of the interface representation. The VOF method implicitly represents the interface between two immiscible fluids $a$ and $b$ in each computational cell $\Omega_{i}$ as a local fraction of volume
\begin{equation}
    \alpha=\frac{1}{\mathcal{V}_{\Omega_{i}}}\int_{\Omega_{i}}\chi(\mathbi{x})dV,
\end{equation}
where $\mathcal{V}_{\Omega_{i}}$ is the cell volume, and $\chi$ is an indicator function that follows
\begin{equation}
    \chi(\mathbi{x})=
    \begin{cases}
    1 & \text{for } \mathbi{x}\in \text{fluid } a \\
    0 & \text{for } \mathbi{x}\in \text{fluid } b.
    \end{cases}
\end{equation}
The mean curvature, $H$, can be approximately computed from $\alpha$ by calculating the surface normal $\mathbi{n}=-\nabla\alpha/\left|\nabla\alpha\right|$ and its divergence $2H=-\nabla\cdot\mathbi{n}$, where the surface normal points from fluid $a$ to fluid $b$, although this calculation can be improved by smoothing the discontinuous volume fraction field through convolution with a kernel function \cite{Brackbill1992}. 
The reconstructed distance function (RDF) method forms a smooth level set distance function from the volume fractions, but neither the RDF or convolution-based methods converge with mesh refinement \cite{Cummins2005EstimatingFractions}. \par
A popular approach to curvature estimation is the height function (HF) method \cite{Nichols1980SOLA-VOF:Boundaries,Hirt1981,Torrey1985NASA-VOF2D:Surfaces}, which analytically \cite{Bornia2011} and numerically \cite{Cummins2005EstimatingFractions} converges with mesh refinement. The height function method integrates the volume fraction field along columns of a Cartesian mesh to form a stencil of heights, after which the curvature is computed using finite difference operators on those heights. While the original formulation of the HF method is second-order accurate, the method was extended with fourth-order \cite{Bornia2011,Sussman2007,Francois2010InterfaceGrids} and then arbitrary-order \cite{Zhang2017HFES:Curves,Evrard2020} formulations in two and three dimensions. Moreover, while the original HF formulation requires a uniform Cartesian mesh, the HF method has been extended to non-uniform Cartesian meshes \cite{Francois2010InterfaceGrids,Evrard2020}. The HF method has also been extended to adaptively refined meshes \cite{Popinet2009} and unstructured meshes \cite{Ivey2015}, in both cases by projecting the underlying volume fraction field onto a uniform, Cartesian stencil and interpolating the volume fraction data to reconstruct volume fractions on the Cartesian stencil.
The HF method has also been coupled with other methods to compute curvatures where the local interface is highly curved and a consistent stencil of heights cannot be formed \cite{Popinet2009,Owkes2015,Patel2018ComputingApproach,Karnakov2020}.

Another curvature estimation approach is to utilize the piecewise linear interface calculation (PLIC) reconstructions \cite{DeBar1974FundamentalsCode,Youngs1982Time-dependentDistortion}. A paraboloid can be fitted to the centroids of the PLIC reconstruction polygons (henceforth referred to as ``PLIC centroids") by solution of a least-squares problem, from which the curvature can be directly computed \cite{Scardovelli2003InterfaceAdvection,Popinet2009,Owkes2018}. Several techniques for computing curvatures from pointwise data can also be found in the computer vision literature \cite{Taubin1991EstimationSegmentation,Goldfeather,Magid2007}. In contrast to methods that utilize the PLIC centroids, the method of Jibben et al. \cite{Jibben2019} fits a paraboloid to the PLIC surface in a volumetric manner. It forms height-like columns from the projection of the PLIC reconstruction polygons onto a reference plane and chooses the paraboloid that best approximates the volume of the columns. 
A paraboloid can also be fitted to a neighborhood of interfacial cells by matching, either exactly or in a least-squares manner, the intersection volume of the paraboloid and each interfacial cell to the cell volume fraction. The  parabolic reconstruction of surface tension (PROST) method \cite{Renardy2002PROST:Method} fits an implicitly defined paraboloid onto volume fractions in a 3D Cartesian mesh. For the integration of the paraboloid-cell intersection volume, it uses an approximation that yields second-order accurate volumes. The method of Evrard et al. \cite{Evrard2017a} exactly fits a parabola to the volume fractions of three cells in a 2D unstructured mesh. It achieves the same order of accuracy as the height function method, even for fine meshes. A curvature estimation method utilizing exact volume integration for 3D meshes does not exist in the prior literature, likely due to the computational expense and/or complexity of the volume moments calculation in three dimensions. Furthermore, the methods utilizing cell-paraboloid intersection volumes require expensive non-linear optimization of the fitted paraboloid coefficients.
Finally, machine learning has been used to predict mean curvature directly from volume fractions for Cartesian meshes \cite{Meier2002AMethods,Qi2019ComputingLearning,Patel2019ComputingApproach}. The curvature errors from machine learning methods, however, have not been shown to converge with mesh refinement.

This work compares the computational cost and error convergence of four curvature evaluation methods: (i) the height function method, (ii) the PLIC-centroidal paraboloid fitting method, (iii) the PLIC-volumetric fitting method of Jibben et al. \cite{Jibben2019}, and (iv) a novel direct volumetric fitting method in three dimensions. This work refers to the methods by the abbreviations HF, PC, PV, and VF, respectively. Section \ref{methods} details the mathematical operations performed in each method. Section \ref{sources_of_error} discusses sources of curvature estimation error and the expected mesh convergence behavior of curvature estimation error. The four methods are first compared using a series of randomized paraboloids as reference interfaces in Section \ref{static_tests}. Next, Section \ref{dynamic_tests} examines the performance of the methods when the curvature calculation is coupled to a two-phase Navier-Stokes solver in stationary and translating droplet test cases. Finally, conclusions are drawn in Section \ref{conclusions}.

\section{Curvature evaluation methods}\label{methods}
\subsection{Height function}\label{heightfcn}
The height function method integrates the volume fractions of the target cell and neighboring cells along columns in the pseudo-normal direction of the target cell to form a stencil of heights. The pseudo-normal direction is the Cartesian direction $x,y,$ or $ z$ with the largest absolute component of the interface normal vector. For a target cell with index $(i_c,j_c,k_c)$, assuming that the pseudo-normal direction is $z$, the heights $h$ are computed as
\begin{equation}
    h_{i,j}=\sum_{k=k_c-(N_H-1)/2}^{k_c+(N_H-1)/2}\alpha_{i,j,k}\Delta z\quad\text{for}\quad
    \begin{cases}
    i_c-(N_N-1)/2\le i \le i_c+(N_N-1)/2 \\
    j_c-(N_N-1)/2\le j \le j_c+(N_N-1)/2,
    \end{cases}
\end{equation} 
where $N_H$ is the number of cells in each column, while $N_N$ is the width of the height stencil. To obtain accurate interfacial heights, the column height $N_H$ must be large enough such that the column contains at least one cell with $\alpha=0$ and at least one cell with $\alpha=1$. 
For a second-order HF method, the first and second partial derivatives are computed from the heights using central differences as 
\begin{subequations}
    \begin{align}
        h_x&\approx\frac{h_{i+1,j}-h_{i-1,j}}{2\Delta x}, \\
        h_y&\approx\frac{h_{i,j+1}-h_{i,j-1}}{2\Delta y}j, \\
        h_{xx}&\approx\frac{h_{i+1,j}-2h_{i,j}+h_{i-1,j}}{\Delta x^2}, \\
        h_{yy}&\approx\frac{h_{i,j+1}-2h_{i,j}+h_{i,j-1}}{\Delta y^2}, \\
        h_{xy}&\approx\frac{h_{i+1,j+1}-h_{i-1,j+1}-h_{i+1,j-1}+h_{i-1,j-1}}{2\Delta x \,2 \Delta y}.
    \end{align}
\end{subequations}
The mean curvature in the target cell can then be estimated from the partial derivatives with
\begin{equation}\label{H}
    H=-\frac{h_{xx}+h_{yy}+h_{xx}h_y^2+h_{yy}h_x^2-2h_{xy}h_xh_y}{2(1+h_x^2+h_y^2)^{3/2}}.
\end{equation}
\subsection{PLIC-centroidal fitting}
The PLIC-centroidal method fits a paraboloid to a neighborhood of PLIC centroids inside of and surrounding the target cell by using a weighted least-squares regression. The paraboloid is defined in a coordinate system $(x',y',z')$, where $z'$ is aligned with the target cell's interface normal vector $\mathbf{\hat n}$, and the target cell's PLIC centroid is located at $(x',y',z')=(0,0,0)$. The paraboloid takes the form
\begin{equation}\label{fit_paraboloid}
    z'=f(x',y')=a_0+a_1x'+a_2y'+a_3x'^{2}+a_4x'y'+a_5y'^{2},
\end{equation}
such that the paraboloid axis is parallel to $\mathbf{\hat n}$. 

Let $\mathbi{x}_i^{\prime}=(x'_i,y'_i,z'_i)$ be the centroid of a PLIC reconstruction polygon with index $i$ in the neighborhood $\calN_c$ of the target cell.
The paraboloid coefficients minimize the cost function
\begin{equation}\label{J_centroid}
    J(\mathbi{a})=\sum_{i\in\calN_c} w^R_i\left(w^A_i\left(z'_i-f(x_i^{\prime},y'_i)\right)\right)^2,
\end{equation}
where $\mathbi{a}=[a_0,a_1,a_2,a_3,a_4,a_5]$, and $w^R$ and $w^A$ are weights associated with each centroid. Minimizing $J$ is equivalent to minimizing the algebraic distance between the centroids and the paraboloid. This is different than minimizing the Euclidean distance between the centroids and the paraboloid, but it enables the formulation of the minimization problem as a linear system.
The radial distance-based weight $w^R$ is given by the Wendland radial basis function \cite{Wendland1995PiecewiseDegree}
\begin{equation}\label{w_R}
    w^R_i=
    \begin{cases}
    (1+4r/d)(1-r/d)^4 & \text{for } 0 \le r \le d \\ 
                    0 & \text{for } r > d,
    \end{cases}
\end{equation}
where $r=\Vert\mathbi{x}_i^{\prime}\Vert$, and $d$ is the width of the weighting function. Since the paraboloid fit is, in essence, a second-order Taylor series approximation of the interface at a given point $\mathbi{x}$, its validity decreases with increasing distance from $\mathbi{x}$ for non-paraboloid interfaces. Therefore, the distance-based weighting is necessary to maintain the locality of the fit and prevent outlier points from strongly influencing the fit. The distance-based weighting is similar to those utilized in the PROST method \cite{Renardy2002PROST:Method} and the Adjustable Curvature Evaluation Scale (ACES) method of Owkes et al. \cite{Owkes2018}. As shown in Section \ref{dynamic_tests}, the radial weighting greatly influences the production of spurious velocities that result from the coupling of the curvature calculation with a two-phase Navier-Stokes solver.
The surface area weight $w^A$ is given by
\begin{equation}
    w^A_i=\mathcal{A}_i(\mathbf{\hat n}\cdot\mathbf{\hat n}_i),
\end{equation}
where $\mathcal{A}$ is the area of the PLIC reconstruction polygon. The term $\mathcal{A}_i(\mathbf{\hat n}\cdot\mathbf{\hat n}_i)$ is the area of the PLIC reconstruction polygon projected onto the $(x',y')$ plane. The products $\mathcal{A}_i(\mathbf{\hat n}\cdot\mathbf{\hat n}_i)z'_i$ and $\mathcal{A}_i(\mathbf{\hat n}\cdot\mathbf{\hat n}_i)f(\mathbi{x}_i^{\prime})$ are therefore approximations of the volumes underneath the PLIC reconstruction and the fitted paraboloid, respectively, and the area-weighted centroid fitting method therefore approximates a volume-matching method.


Differentiation of Eq.\ \eqref{J_centroid} with respect to $\mathbi{a}$ results in a $6\times6$ linear system which can be solved for $\mathbi{a}$.
The mean curvature can be computed directly on the paraboloid at $(x',y')=(0,0)$ as
\begin{equation}\label{H-centroid}
    H=-\frac{a_{5}a_{1}^{2}-a_{4}a_{1}a_{2}+a_3a_2^2+a_3+a_5}{\left(1+a_1^2+a_2^2\right)^{3/2}}.
\end{equation}
\subsection{PLIC-volumetric fitting}
The PLIC-volumetric fitting method fits a paraboloid to a neighborhood of PLIC reconstruction polygons such that the volume underneath the paraboloid matches the volume underneath the interface polygons in a least-squares sense. As in the PLIC-centroidal method, the paraboloid is defined in a coordinate system $(x',y',z')$, where $z'$ is aligned with the target cell's interface normal vector $\mathbf{\hat n}$, and the target cell's PLIC centroid is located at $(x',y',z')=(0,0,0)$. The method chooses the paraboloid in the form of Eq.\ \eqref{fit_paraboloid} with coefficients $\mathbi{a}$ that minimizes the cost function
\begin{equation}\label{jibben_costfcn}
    J(\mathbi{a})=\sum_{p}w^R_p\left( \int_{\Gamma_p} \left(f(x',y')-g_p(x',y')\right) dA \right)^2,
\end{equation}
where $w^R$ is the distance-based weight given by Eq.\ \eqref{w_R}, the domain $\Gamma_p$ is the projection of the interface polygon of index $p$ onto the $(x',y')$ plane, and
\begin{equation}
	g_p(x',y')=b_{p,0}+b_{p,1}x'+b_{p,2}y'
\end{equation}
is the plane containing the interface polygon. While the original method of Jibben et al. \cite{Jibben2019} uses a uniform weight $w^R=1$, the distance-based weight from Eq.\ \eqref{w_R} is used here to localize the fit as in the PLIC-centroidal method. For a polygon $p$, let  $\left(x'_{p,v},y'_{p,v}\right)$ represent the $\left(x',y'\right)$ coordinates of a vertex of index $v$, where $v$ increases in the counter-clockwise direction with respect to the plane normal vector. 

Minimizing Eq.\ \eqref{jibben_costfcn} with respect to the coefficients $\mathbi{a}$ results in the set of equations
\begin{equation}\label{dJ}
    \sum_{p}w^R_p\left( \int_{\Gamma_p} \left(f(x',y')-g_p(x',y')\right) dA \right)\left( \int_{\Gamma_p} \phi dA \right)=0,
\end{equation}
to be solved for each $\phi\in\left\{1,x',y',x'^2,x'y',y'^2\right\}$.
Equation \eqref{dJ} requires the integration of the monomial terms in $f(x',y')$ and $\phi$ within each projected polygon. Using Green's Theorem, the double integrals convert into piecewise line integrals along the perimeter of each projected polygon:
\begin{subequations}
    \begin{align}
        s_{p,0}&= \int_{\Gamma_p}     dA = \frac{1}{2} \sum_{v=1}^{N_{v,p}} \left(x'_{p,v}y'_{p,v+1}-x'_{p,v+1}y'_{p,v}\right),\\
        s_{p,1}&= \int_{\Gamma_p} x'  dA = \frac{1}{6} \sum_{v=1}^{N_{v,p}} \left(x'_{p,v}+x'_{p,v+1}\right)\left(x'_{p,v}y'_{p,v+1}-x'_{p,v+1}y'_{p,v}\right),\\
        s_{p,2}&= \int_{\Gamma_p} y'  dA = \frac{1}{6} \sum_{v=1}^{N_{v,p}} \left(y'_{p,v}+y'_{p,v+1}\right)\left(x'_{p,v}y'_{p,v+1}-x'_{p,v+1}y'_{p,v}\right),\\
        s_{p,3}&= \int_{\Gamma_p} x'^2dA = \frac{1}{12} \sum_{v=1}^{N_{v,p}} \left(x'_{p,v}+x'_{p,v+1}\right)\left(x'^2_{p,v}+x'^2_{p,v+1}\right)\left(y'_{p,v+1}-y'_{p,v}\right),\\
        s_{p,4}&= \int_{\Gamma_p} x'y'dA = \frac{1}{24} \sum_{v=1}^{N_{v,p}} \left(2x'_{p,v}y'_{p,v}+x'_{p,v}y'_{p,v+1}+x'_{p,v+1}y'_{p,v}+2x'_{p,v+1}y'_{p,v+1}\right)\left(x'_{p,v}y'_{p,v+1}-x'_{p,v+1}y'_{p,v}\right), \\ \intertext{and}
        s_{p,5}&= \int_{\Gamma_p} y'^2dA = \frac{1}{12} \sum_{v=1}^{N_{v,p}} \left(y'_{p,v}+y'_{p,v+1}\right)\left(y'^2_{p,v}+y'^2_{p,v+1}\right)\left(x'_{p,v+1}-x'_{p,v}\right),
    \end{align}
\end{subequations}
where $N_{v,p}$ is the number of vertices in polygon $p$. Equation \eqref{dJ} thereby becomes a symmetric linear system of equations for the paraboloid coefficients $\mathbi{a}$,
\begin{equation}
    \begin{gathered}
            \mathbi{A}\mathbi{a}=\mathbi{b}, \\ \text{where} \quad
        A_{ij}=\sum_p w^R_ps_{p,i}s_{p,j} \quad \text{and} \quad b_i=\sum_p w^R_ps_{p,i}\left(b_{p,0}s_{p,0}+b_{p,1}s_{p,1}+b_{p,2}s_{p,2}\right).
    \end{gathered}
\end{equation}
As with the PLIC-centroidal method, the mean curvature can be computed from the partial derivatives of the resulting paraboloid with Eq.\ \eqref{H-centroid}.

\subsection{Volumetric fitting}\label{vf}
The volumetric fitting method fits a paraboloid to a neighborhood of interfacial cells such that the fractions of cell volume underneath the paraboloid match the underlying volume fractions $\alpha$ in a least-squares sense. The method chooses the paraboloid, in the form of Eq.\ \eqref{fit_paraboloid} with coefficients $\mathbi{a}$, that minimizes the cost function
\begin{equation}\label{J-volume}
    J(\mathbi{a})=\sum_{i\in\calN_c}w^R_i\left(\Tilde{\alpha}_i- \alpha_i \right)^2,
\end{equation}
where
\begin{equation}
    \Tilde{\alpha}_i=\frac{1}{\mathcal{V}_{\Omega_i}} \int_{\Omega_i} \tilde{\chi}(x',y',z') dV,
\end{equation}
$w^R_i$ is the distance-based weight given by Eq.\ \eqref{w_R}, the domain $\Omega_i$ is the neighborhood cell with index $i$, $\alpha_i$ is the associated volume fraction, $\mathcal{V}_{\Omega_i}$ is the associated total cell volume, and $\tilde{\chi}$ is an indicator function that follows
\begin{equation}
    \tilde{\chi}(x',y',z')=
    \begin{cases}
    1 & \text{if } z' \le f(x',y') \\
    0 & \text{if } z' > f(x',y').
    \end{cases}
\end{equation}
The intersection volume of the fitted paraboloid $f(x',y')$ and polyhedral interfacial cell $\Omega$ is calculated by successive application of the divergence theorem \cite{Evrard2022}. The analytical integration produces volume fraction errors on the order of double precision machine epsilon at a computational cost several orders of magnitude lower than brute-force numerical integration.

While the minimization problems in the PLIC-centroidal and PLIC-volumetric methods can be formulated into directly-solvable linear systems,
the minimization of Eq.\ \eqref{J-volume} requires an iterative method due to the nonlinear relationship between $\mathbi{a}$ and the intersection volume. 
The minimization of $J$ is performed using the Levenberg-Marquardt algorithm \cite{Levenberg1944ASquares,Marquardt1963AnParameters}, which is a gradient-based local minimization method for solving nonlinear least-squares problems. It requires an initial guess for the paraboloid coefficients $\mathbi{a}$, which is chosen to be the solution from the PLIC-volumetric fitting method. The Jacobian matrix $\mathbf{J}_{ij}=\partial \Tilde{\alpha}_i/\partial a_j$, used to determine the step size and direction, is calculated analytically during each iteration. 

As with the PLIC-centroidal and PLIC-volumetric method, the mean curvature can be computed from the partial derivatives of the resulting paraboloid with Eq.\ \eqref{H-centroid}.
\section{Sources of curvature error}\label{sources_of_error}
Previous work has shown that the curvature error increases when the underlying volume fractions have incurred errors, such as those incurred during the VOF advection step \cite{Zhang2017HFES:Curves,Owkes2018,Remmerswaal2022}. Lemma 3 from the analysis of Remmerswaal and Veldman \cite{Remmerswaal2022} provides an error estimate for the derivatives of the height function when the heights are constructed from inexact volume fractions. Let $f$ be the explicit, local representation of the phase interface. If the height function $h$ is constructed from $p^{\text{th}}$-order volume fractions, the $\lambda^{\text{th}}$ derivative of $h$ can be related to $f$ as
\begin{equation}\label{convergence_order}
    h^{[\lambda]}=f^{[\lambda]}+\order{\Delta^2}+\order{\Delta^{p+1-\lambda}},
\end{equation}
where $\Delta$ is the characteristic mesh size. Since the curvature calculation in Eq.\ \eqref{H} uses the second derivative, the curvature order of accuracy should be $\min(2,p-1)$ when computed with the height function method. While Zhang \cite{Zhang2013OnMethods} shows that standard VOF advection schemes using piecewise linear interface approximations are second-order in the $L_1$ norm of the volume error,  Remmerswaal and Veldman \cite{Remmerswaal2022} show that the accuracy of the volume fractions is first-order in the $\linf$ norm. Consequently, the curvature error becomes zeroth-order. However, the curvature calculation method and its associated parameters can influence the leading-order behavior such that the error still converges within a limited range of mesh sizes \cite{Zhang2017HFES:Curves}. The work of Evrard et al. \cite{Evrard2017a} demonstrates that the height function method in two dimensions is equivalent to a volumetric fitting method that utilizes a parabolic reconstruction. Therefore, Eq.\ \eqref{convergence_order} provides an approximate upper bound for the order of accuracy of a volumetric fitting curvature method and other parabolic fitting methods such as PC and PV. 
\section{Static tests}\label{static_tests}
\subsection{Random paraboloids}
This study compares the curvature evaluation methods from Section \ref{methods} by analyzing the convergence of their respective curvature errors. Unlike previous curvature studies that examine mesh convergence for a limited number of exact shapes with fixed curvature, this study varies the shape curvature while maintaining a fixed mesh size. A characteristic mesh size is first defined as  $\Delta=\left(\Delta x \Delta y \Delta z\right)^{1/3}$. When the curvature is nondimensionalized with the characteristic mesh size as $H\Delta$, the fixed curvature and fixed mesh size approaches are equivalent. The reference interfaces in this study are a series of $N_p=10^6$ randomly generated paraboloids of the form
\begin{equation}
    z'=f(x',y')=A+\beta Bx'+\beta Cy'+\beta Dx'^2+\beta Ex'y'+\beta Fy'^2,
\end{equation}
where
\begin{subequations}
    \begin{align}
          A&\sim\calU[-\sqrt{3}/2,\sqrt{3}/2) \\
        B,C&\sim\calU[-\sqrt{3}  ,\sqrt{3}  ) \\
        D,F&\sim\calU[-2.5       ,2.5       ) \\
          E&=0                                \\
      \beta&=2^{-p}                           \\
          p&\sim\calU[0,28),
    \end{align}    
\end{subequations}
and $\calU[a,b)$ is a uniform distribution over a half-open interval with bounds $a$ and $b$.
The paraboloid coordinate system $(x',y',z')$ is randomly oriented relative to the computational mesh, which allows the coefficient $E$ to be set to zero, as $E$ effectively rotates the paraboloid about the $z'$ axis. The use of paraboloids as reference interfaces better isolates the error associated with the fitting procedure from the error associated with the choice of fitting shape. 
The mesh is an $11\times11\times11$ Cartesian grid with a constant mesh spacing of $\Delta x=\Delta y=\Delta z=\Delta=1$. The origin of the paraboloid coordinate system is placed at the centroid of the centermost cell in the domain. The volume fraction field is initialized using the analytical paraboloid-polyhedron intersection volume moments calculation described in Section \ref{vf}. The PLIC reconstruction in each cell is formed using the LVIRA method \cite{Pilliod2004}. 


As stated in Section \ref{heightfcn}, in the HF method, $N_H$ is the number of cells in each column, while $N_N$ is the width of the height stencil. In this work, $N_H=11$, while $N_N=3$, which corresponds to a second-order HF formulation. A column height of $N_H=11$ is necessary to ensure well-defined interface heights for each random paraboloid given that the interface is sufficiently resolved by the mesh. A fourth-order HF method with $N_N=5$ would require an even larger column height and is therefore not considered in this study. For the PC, PV, and VF methods, the cell neighborhood size is $S\times S\times S$, where three values of the stencil width $S$ are compared, $S=3$, 5, and 7, to evaluate the effect of stencil size on the curvature accuracy. Since the fitted shape and the reference interface are both paraboloids, the locality of the fit is not an issue in this test, and therefore, the radial weighting for the fitting methods is uniform, i.e., $w^R=1$.

The curvature errors are divided into 11 bins according to the exact curvature of the reference paraboloid, where the bin edges are logarithmically spaced such that there is an approximately equal number of paraboloids represented in each bin. The exact curvature of each randomly chosen paraboloid is computed by projecting the PLIC centroid onto the reference paraboloid in the direction of the PLIC interface normal and evaluating the curvature at the projected centroid location. For the paraboloids within each bin, two error norms are computed that compare the exact curvature $H_e$ to the estimated curvature $H$. They are defined as 
\begin{align}
    \ltwo(H_{e})&\equiv \frac{\sqrt{\sum_{i=1}^{N_b}\left(H_i-H_{e,i}\right)^2}}{\sqrt{\sum_{i=1}^{N_b} C_{e,i}^2}} \\
    \linf(H_{e})&\equiv \max_{i\in\{1,...,N_b\}} \left|\frac{H_i-H_{e,i}}{C_{e,i}}\right|,
\end{align}
where $N_b$ is the number of random paraboloids in each bin, and $C_e$ is the curvedness  \cite{Koenderink1990SolidShape} of the reference paraboloid, calculated at the same location as $H_e$. 
The curvedness at a point on a surface is
\begin{equation}
    C=\sqrt{\frac{\kappa_{1}^2+\kappa_{2}^2}{2}},
\end{equation}
where $\kappa_1$ and $\kappa_2$ are the principal curvatures at the location of interest. Curvedness is a shape-independent metric of the ``intensity" of curvature, and normalizing the mean curvature error by the curvedness allows for the error convergence analysis of curvatures calculated on arbitrarily-shaped paraboloids as specified by their coefficients. 
Curvedness can also be derived from the mean and Gaussian curvatures using
\begin{equation}
    C=\sqrt{2H^2-K},
\end{equation}
where the Gaussian curvature is
\begin{equation}
    K=\frac{f_{xx}f_{yy}-f_{xy}^2}{\left(1+f_x^2+f_y^2\right)^2}.
\end{equation}
As the HF method can fail for certain paraboloids with large curvatures where a consistent stencil of heights cannot be formed, the value of $N_b$ is reduced in the corresponding bins, only in the calculation of error norms for the HF method. In the random paraboloid test cases, the HF method with $N_H=11$ fails for about 51\% of paraboloids with $C>10^{-1}$, 0.066\% of paraboloids with $10^{-1}\ge C \ge 10^{-2}$, and zero paraboloids below $C\le 10^{-2}$.

\begin{figure}[p]
    \centering
    \includegraphics[width=1\columnwidth]{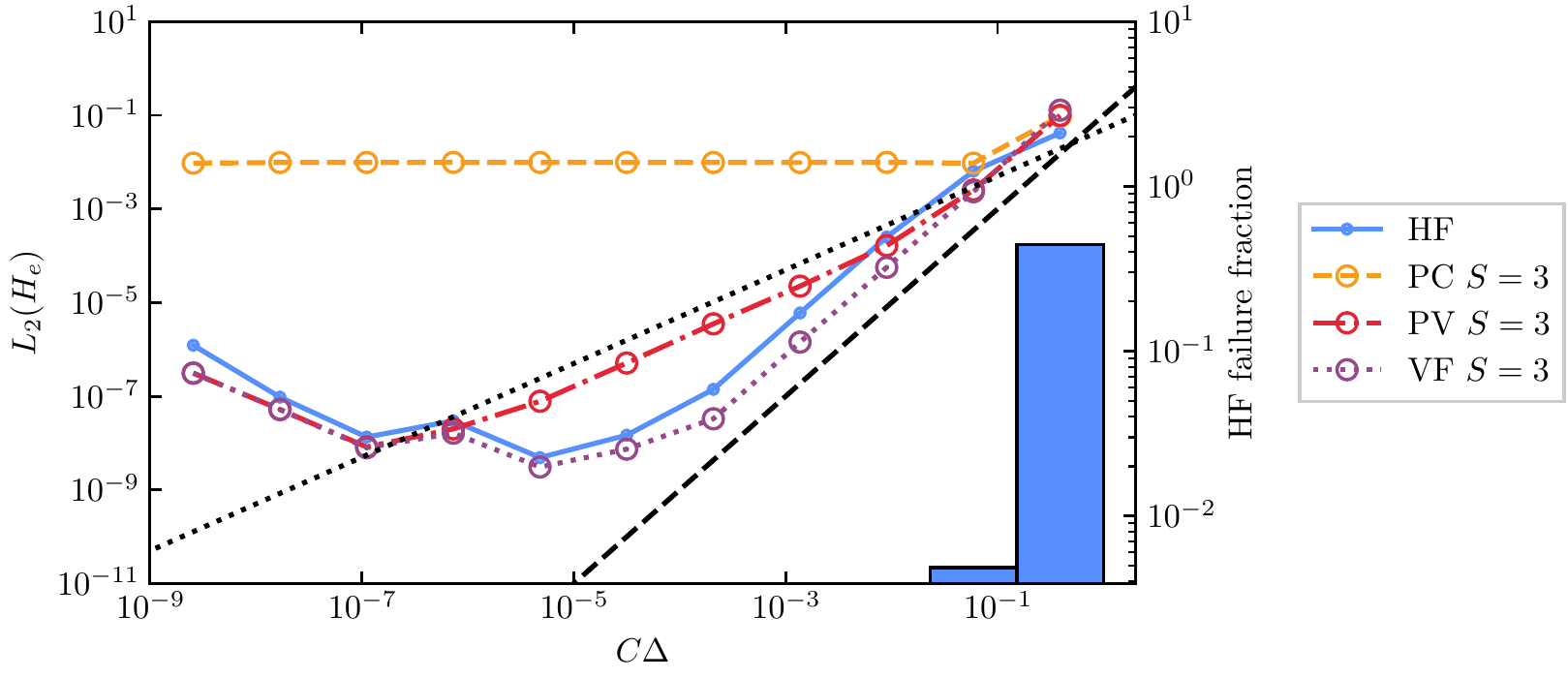}
    \caption{$L_2$ error norm of the curvature error as a function of nondimensional paraboloid curvature resulting from the utilization of the height function (HF), PLIC-centroidal (PC), PLIC-volumetric (PV), and volumetric fitting (VF) methods on randomly chosen paraboloids. For the PC, PV, and VF methods, the stencil width $S=3$. The dotted and dashed lines represent first- and second-order convergence, respectively. The fraction of paraboloids for which the HF method fails is overlaid as a histogram.}
    \label{fig:random-parab-semi-vof-l2}
\end{figure}

\begin{figure}[p]
    \centering
    \includegraphics[width=1\columnwidth]{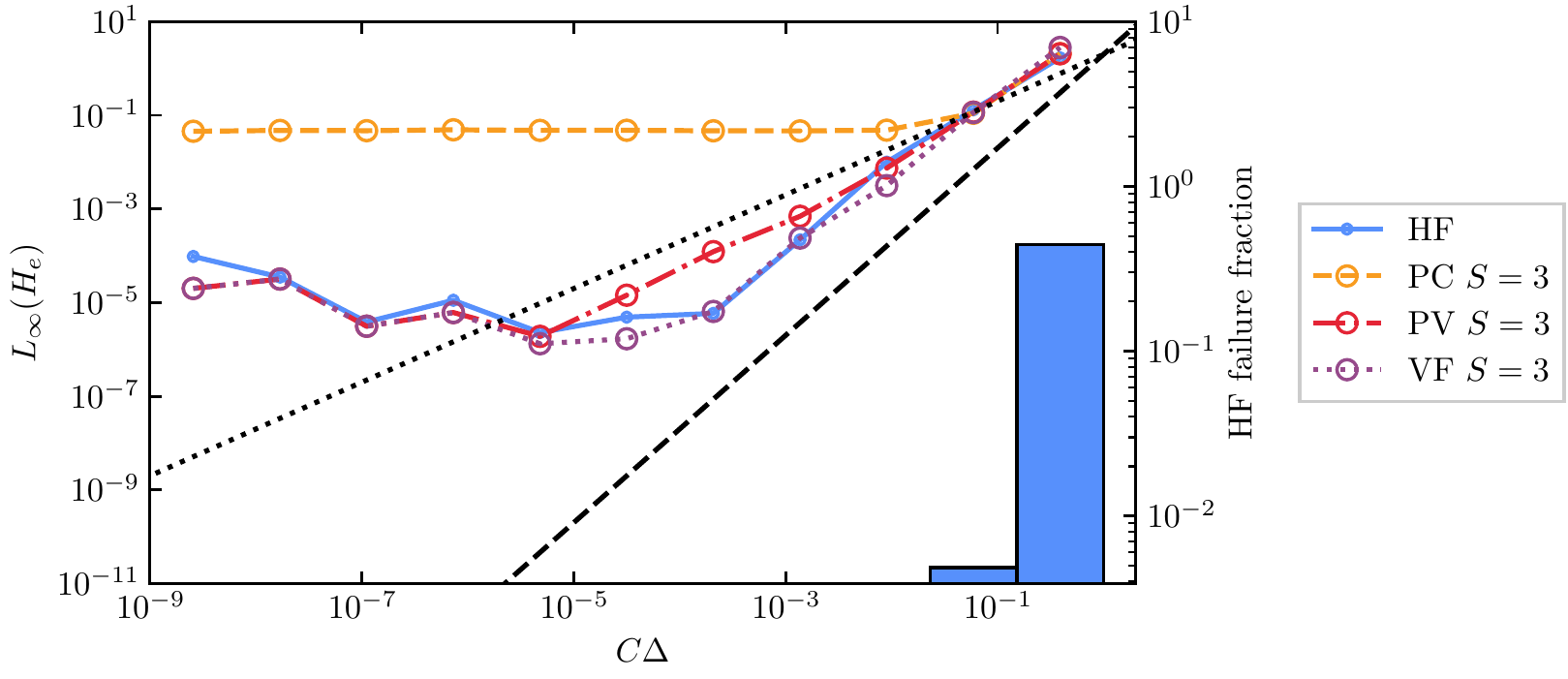}
    \caption{$L_\infty$ error norm of the curvature error as a function of nondimensional paraboloid curvature resulting from the utilization of the height function (HF), PLIC-centroidal (PC), PLIC-volumetric (PV), and volumetric fitting (VF) methods on randomly chosen paraboloids. For the PC, PV, and VF methods, the stencil width $S=3$. The dotted and dashed lines represent first- and second-order convergence, respectively. The count of paraboloids for which the HF method fails is overlaid as a histogram.}
    \label{fig:random-parab-semi-vof-linf}
\end{figure}

Figures \ref{fig:random-parab-semi-vof-l2} and \ref{fig:random-parab-semi-vof-linf} respectively show the $\ltwo$ and $\linf$ norms of the curvature errors for the randomly chosen paraboloids, where $S=3$ for the PC, PV, and VF methods. The HF method demonstrates second-order accuracy as expected from the analytical result. The HF error stops converging at $C\Delta \approx 10^{-5}$, where the unnormalized curvature error is about $10^{-10} \times 10^{-5} = 10^{-15}$, which is near the double precision limit of $\epsilon_{\text{mach}} = 2^{-52} \approx 2\times 10^{-16}$ used in this work. The VF method demonstrates the same second-order convergence as the HF method. The PV method converges with second-order accuracy for $C\Delta>10^{-2}$, but decays entirely to first-order accuracy for $C\Delta<10^{-2}$. The curvature errors of the PC method fail to converge for curvatures $C\Delta<10^{-2}$. For reference, a spherical droplet with $C\Delta=10^{-2}$ is already resolved by $D/\Delta=200$ cells per diameter. Overall, especially when accounting for the magnitude of the curvature error, there is a negligible difference in performance between the PV, HF, and VF methods at typical levels of resolution and a more noticable difference for extremely well resolved interfaces.

To assess the effect of volume fraction errors on the curvature evaluation accuracy, random perturbations are added to the volume fractions of interfacial cells such that
\begin{equation}
    \tilde{\alpha}=\alpha+\Delta \alpha.
\end{equation}
The perturbations $\Delta \alpha$ are sampled from a uniform distribution $\Delta \alpha \sim\calU[-kC,kC)$
whose bounds are determined by the exact curvedness $C$ of the randomly generated paraboloid and an arbitrary coefficient $k$, for which values of $k=0.01$ and $0.1$ are compared in this study. The scaling of the perturbations by $C$ causes the volume fractions to be first-order accurate with respect to the non-dimensional length $C\Delta$, which mimics the accuracy of volume fractions when using a standard second-order accurate VOF transport method based on semi-Lagrangian remapping \cite{Owkes2014}. The perturbed volume fractions are clipped such that $0\le \tilde{\alpha} \le 1$.

\begin{figure}[p]
    \centering
    \includegraphics[width=1\columnwidth]{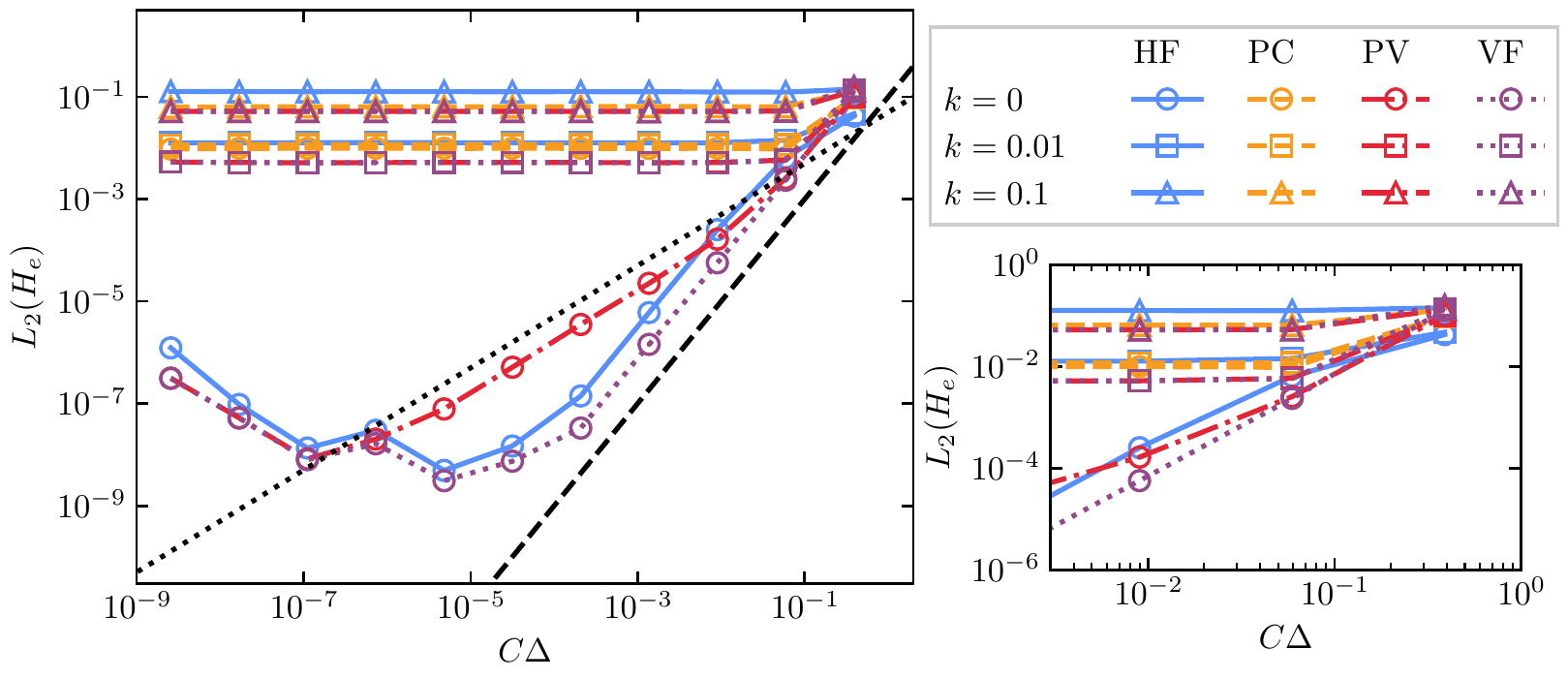}
    \caption{Effect of volume fraction error on the $L_2$ error norm of the curvature error as a function of nondimensional paraboloid curvature resulting from the utilization of the height function (HF), PLIC-centroidal (PC), PLIC-volumetric (PV), and volumetric fitting (VF) methods on randomly chosen paraboloids. The prescribed volume fraction errors are first-order with respect to nondimensional paraboloid curvature and proportional to the coefficient $k$. For the PC, PV, and VF methods, the stencil width $S=3$. The dotted and dashed lines represent first- and second-order convergence, respectively.}
    \label{fig:random-parab-vof-error-l2}
\end{figure}

\begin{figure}[p]
    \centering
    \includegraphics[width=1\columnwidth]{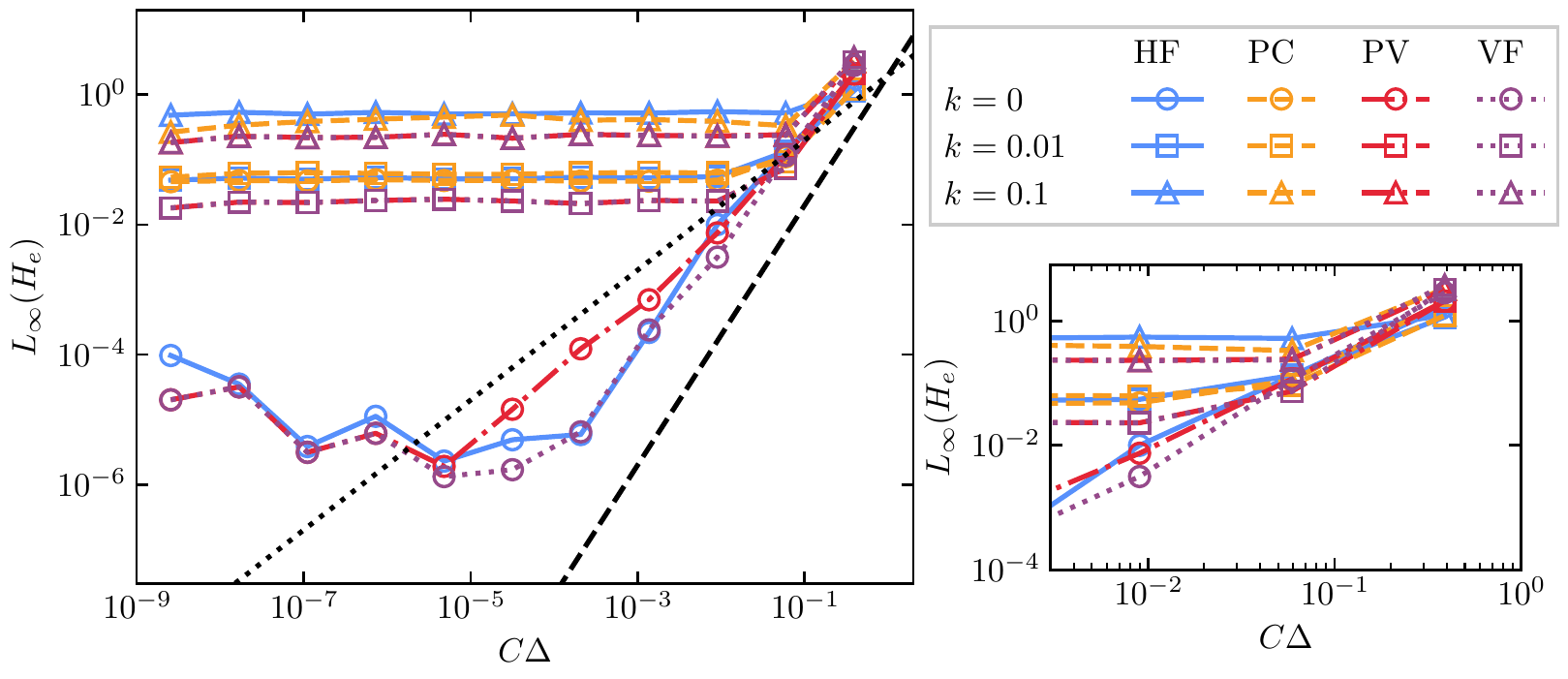}
    \caption{Effect of volume fraction error on the $L_\infty$ error norm of the curvature error as a function of nondimensional paraboloid curvature resulting from the utilization of the height function (HF), PLIC-centroidal (PC), PLIC-volumetric (PV), and volumetric fitting (VF) methods on randomly chosen paraboloids. The prescribed volume fraction errors are first-order with respect to nondimensional paraboloid curvature and proportional to the coefficient $k$. For the PC, PV, and VF methods, the stencil width $S=3$. The dotted and dashed lines represent first- and second-order convergence, respectively.}
    \label{fig:random-parab-vof-error-linf}
\end{figure}

The results of the $k=0.01$ and $0.1$ perturbations are shown in Figures \ref{fig:random-parab-vof-error-l2} and \ref{fig:random-parab-vof-error-linf}. Even with the lower $k=0.01$ perturbations, the curvature error fails to converge for $C\Delta < 10^{-2}$ using any method. This is consistent with the prediction from Eq.\ \eqref{convergence_order}, where the zeroth-order contribution of the volume fraction error becomes the leading-order term for low enough $C\Delta$. When the volume fraction error is increased by using a $k=0.1$ perturbation, the curvature errors increase and fail to converge for any tested range of $C\Delta$. The overall errors from the optimization-based PC, PV, and VF methods are lower than those from the HF method, with the PV and VF methods producing the lowest errors.

\begin{figure}[p]
    \centering
    \includegraphics[width=1\columnwidth]{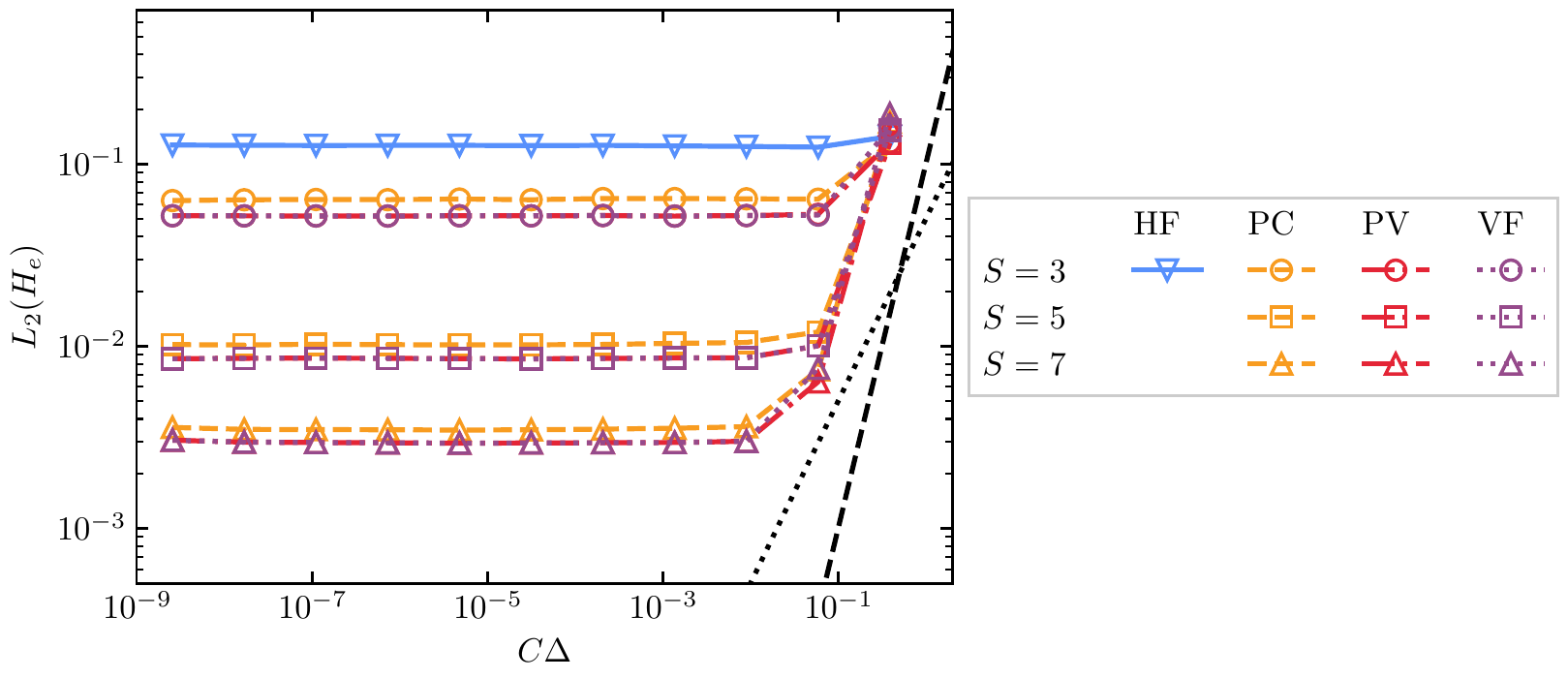}
    \caption{Effect of stencil size on the $L_2$ error norm of the curvature error as a function of nondimensional paraboloid curvature resulting from the utilization of the height function (HF), PLIC-centroidal (PC), PLIC-volumetric (PV), and volumetric fitting (VF) methods on randomly chosen paraboloids. The initialized volume fractions are prescribed a first-order error with respect to nondimensional paraboloid curvature multiplied by a coefficient $k=0.1$. The dotted and dashed lines represent first- and second-order convergence, respectively.}
    \label{fig:random-parab-stencil-l2}
\end{figure}

\begin{figure}[p]
    \centering
    \includegraphics[width=1\columnwidth]{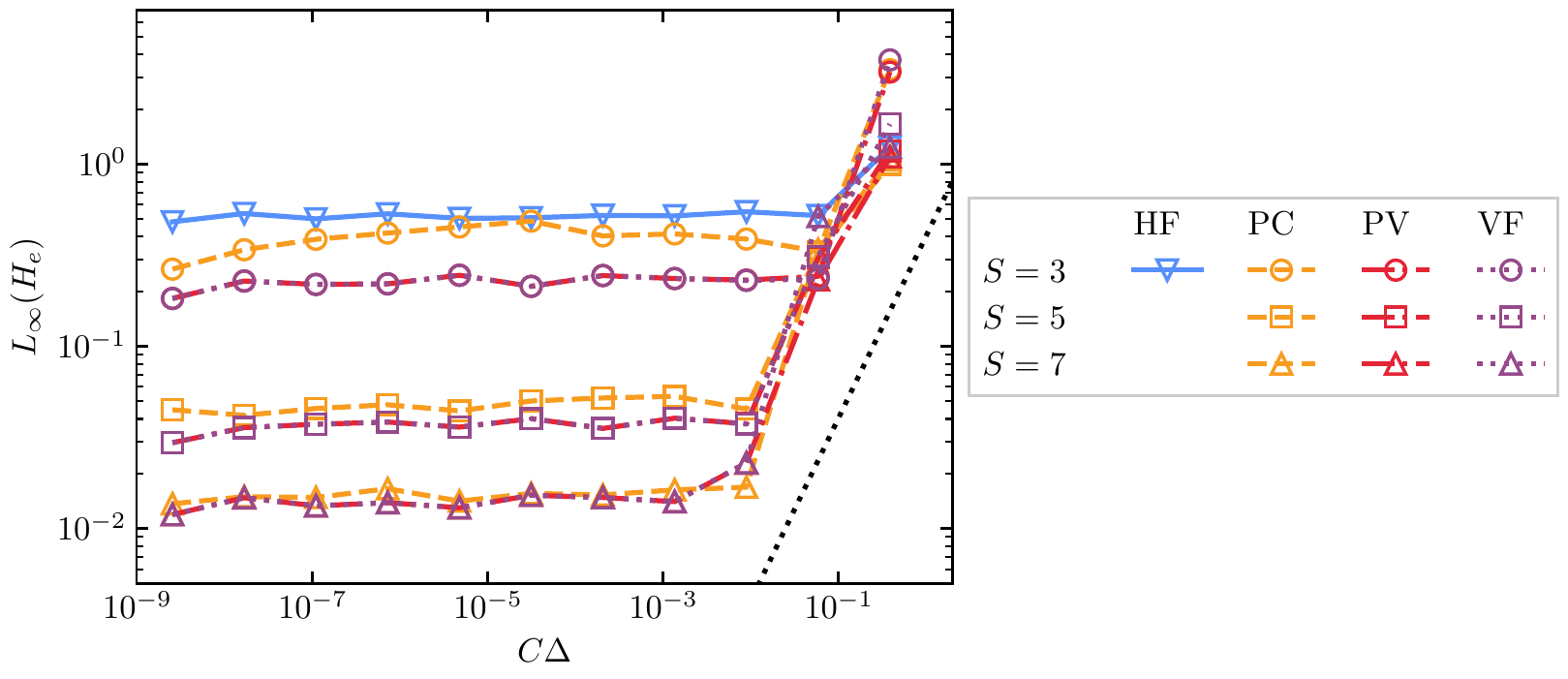}
    \caption{Effect of stencil size on the $L_\infty$ error norm of the curvature error as a function of nondimensional paraboloid curvature resulting from the utilization of the height function (HF), PLIC-centroidal (PC), PLIC-volumetric (PV), and volumetric fitting (VF) methods on randomly chosen paraboloids. The initialized volume fractions are prescribed a first-order error with respect to nondimensional paraboloid curvature multiplied by a coefficient $k=0.1$. The dotted line represents first-order convergence.}
    \label{fig:random-parab-stencil-linf}
\end{figure}

The effect of stencil size is investigated by increasing the stencil size to $S=5$ and $7$ for the PC, PV, and VF methods while maintaining the $k=0.1$ perturbed volume fractions. The effect of stencil size on the accuracy of the HF method is not investigated in this test, and the original HF result is shown for reference. Figures \ref{fig:random-parab-stencil-l2} and \ref{fig:random-parab-stencil-linf} compare the curvature error convergence between the three stencil sizes $S=3,5$, and $7$ for the PC, PV, and VF methods. With each increase in stencil size above $S=3$, the range of $C\Delta$ for which the curvature error converges increases. With a stencil size of $S=5$, the curvature error stops decreasing below $C\Delta = 10^{-1}$. With a stencil size of $S=7$, however, the error decays from approximately second-order to first-order around $C\Delta = 10^{-1}$ and then stops converging below $C\Delta = 10^{-3}$. 

\begin{figure}[tb]
    \centering
    \includegraphics[width=1.0\columnwidth]{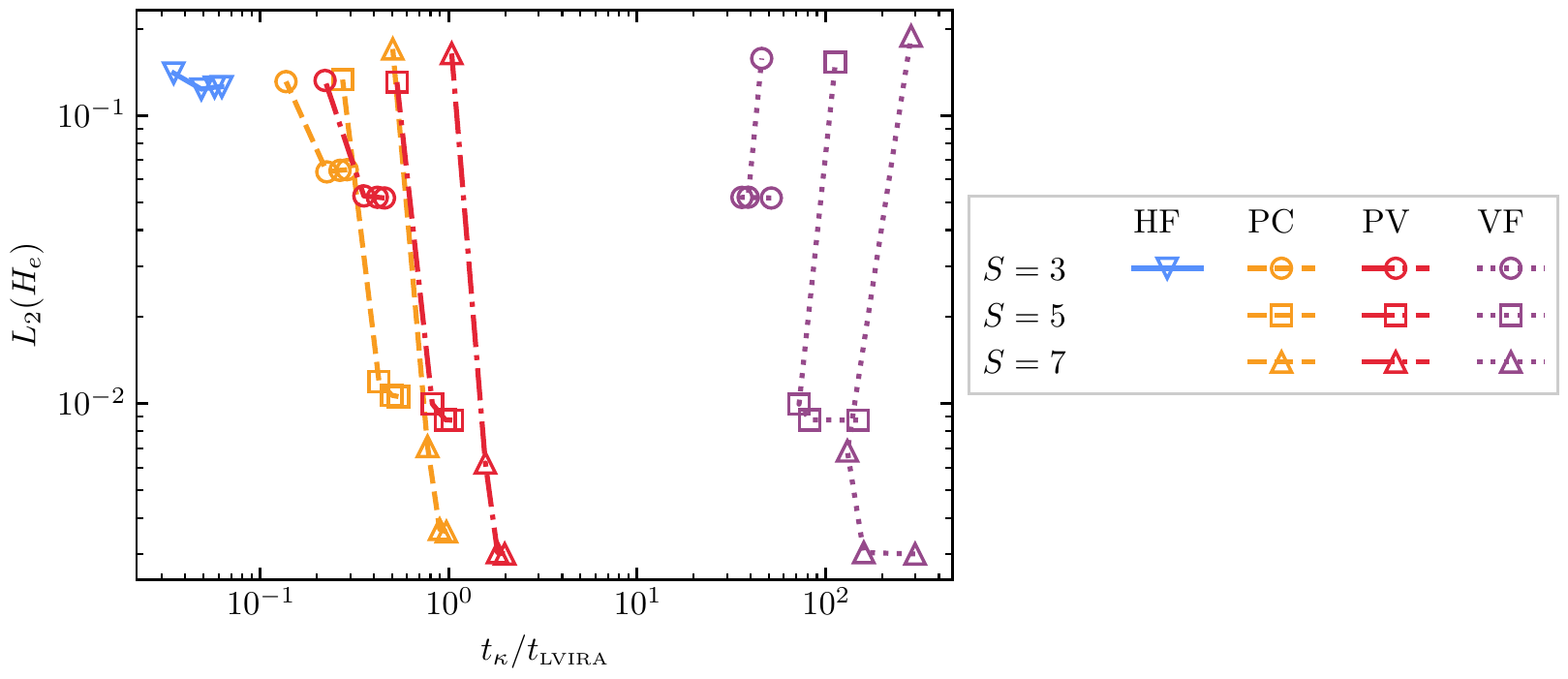}
    \caption{$\ltwo$ norm of the curvature error versus curvature evaluation time for the random paraboloid case. The errors for paraboloids with $C\Delta > 1.5\times 10^{-4}$ are shown. The initialized volume fractions are prescribed a first-order error with respect to nondimensional paraboloid curvature multiplied by a coefficient $k=0.1$.}
    \label{fig:cost}
\end{figure}

Figure \ref{fig:cost} compares the evaluation time and $\ltwo$ curvature error for the tested methods when the curvatures are computed from a $k=0.1$ perturbed volume fraction field. The curvature evaluation time is normalized by the time necessary to perform the LVIRA interface reconstruction for the centermost cell. Results are shown for $C\Delta>5.3\times10^{-4}$, corresponding to the rightmost four bins in Figures \ref{fig:random-parab-semi-vof-l2} through \ref{fig:random-parab-stencil-linf}, as all tested methods do not show error convergence for lower $C\Delta$. Furthermore, the rightmost four bins already represent a very large curvature range, as a spherical droplet with $C\Delta>5.3\times 10^{-4}$ would be resolved by about $D/\Delta=3800$ cells per diameter. The geometric mean of the time is chosen as the representative time for each bin. The timings show that the PC and PV methods, regardless of stencil size, have computational costs that are about one order of magnitude greater than those of the HF method and the same order of magnitude as those of the interface reconstruction. Both the PC and PV methods, when using a $S=3$ stencil, however, result in errors that are about 50\% lower than those of the HF method, with the exception of those of the bin corresponding to the paraboloids with the largest $C\Delta$. In fact, for this bin with $1.5\times 10^{-1} < C\Delta < 1$, the errors are appoximately equal across the methods and stencil sizes with the $S=7$ stencil producing slightly larger errors. Increasing the stencil size to $S=7$ for the PC and PV methods can lower the errors by an over an order of magnitude at two to four times greater cost than with the $S=3$ stencil, while the cost and error associated with an $S=5$ stencil are in between those of the $S=3$ and $S=7$ stencils. The VF method using an $S=3$ stencil has a computational cost that is two orders of magnitude greater than those of the PC and PV methods, but the error is approximately equal to those of the PC and PV methods. In general, the optimization-based methods are more computationally expensive than the HF method but are more robust to errors in the volume fraction field owing to the ability to increase the stencil size at low additional cost. In addition, the additional computational expense of the PC and PV methods over the HF method is insignificant relative to the cost of the interface reconstruction, much less other steps in interfacial flow simulations, such as solving the pressure equation. The PC and PV methods with $S=5$ stencil provide the best balance of computational cost and curvature accuracy, as they provide a large reduction in error over those with the $S=3$ stencil while avoiding the higher errors with the $S=7$ stencil for paraboloids with large $C\Delta$. The subsequent studies therefore utilize them, along with the HF method, to examine the coupling of the curvature calculation with the two-phase Navier–Stokes solver. 

\section{Dynamic tests}\label{dynamic_tests}
The coupling of the curvature calculation to a Navier-Stokes flow solver via the surface tension force is examined with two-dimensional stationary droplet and three-dimensional translating droplet test cases. The two-phase, incompressible Navier-Stokes equations are solved with the NGA2 flow solver \cite{Desjardins2008,Desjardi/NGA2:Solver}. As in the static test case, the PLIC reconstruction in each cell is formed using the LVIRA method \cite{Pilliod2004}. Advection of the volume fraction and momentum fields is performed with the method of Owkes and Desjardins \cite{Owkes2014}, while the surface tension force is calculated with the CSF method \cite{Brackbill1992,Popinet2009}. For the curvature calculation in both test cases, the HF, PC, and PV methods are compared. The HF method is used with a column height of $N_H=7$, which is commonly used in the literature \cite{Cummins2005EstimatingFractions,Francois2006AFramework,Popinet2009,Owkes2018,Jibben2019}, and the PC and PV methods are used with a stencil length of $S=5$ and a radial weighting width of $d=2.5$. 
An additional set of simulations is performed of the translating droplet case using a radial weighting width of $d=3.5$ to examine the effect of weighting on the curvature calculation. As the HF method often fails to estimate curvature for highly curved or ill-resolved interfaces with large $C\Delta$, the PV method is used for cells where the HF method fails.

The volume fractions of interfacial cells are initialized using an octree adaptive mesh refinement, where the original cell is subdivided five times. The marching tetrahedra algorithm \cite{DOI1991AnCells} is then used to calculate volume fractions of the finest subcells and provides second-order accuracy to the initialization. This volume fraction initialization method is similar to those utilized by previous curvature estimation studies \cite{Cummins2005EstimatingFractions,Ivey2015,Owkes2018,Jibben2019,Lopez2009AnFractions}.

\subsection{2D stationary droplet}
A circular droplet initialized in a quiescent flow should remain at rest due to the exact balance between the surface tension force and the pressure jump across the interface. Inaccuracies in the curvature calculation, however, will induce so-called spurious currents that act on the interface to restore numerical balance \cite{Francois2006AFramework,Popinet2009}. This case examines the accuracy of the curvature calculation by measuring the spurious velocities. A circle of diameter $D=0.4$ is placed in a two-dimensional square domain of length $L=1$ with periodic boundary conditions and a uniform Cartesian grid of mesh size $\Delta$. The center of the circle is placed at the center of the domain with a random perturbation in the $x$ and $y$ directions of $\calU[-\Delta/2,\Delta/2)$ to avoid mesh alignment. The viscosity ratio is unity with a viscosity of $\mu=0.1$, and the surface tension coefficient is $\sigma=1$. The density ratio is unity, and the density $\rho$ is the free parameter that modulates the Laplace number $\La=\rho\sigma D/\mu^2$. The spurious velocities are measured with the capillary numbers $\Ca_{\rms}=|\mathbi{u}|_{\rms}\mu/\sigma$ and $\Ca_{\max}=|\mathbi{u}|_{\max}\mu/\sigma$ evaluated at nondimensional time $t/T_\sigma=13.3$, where $|\mathbi{u}|_{\rms}$ is the root mean square velocity, $|\mathbi{u}|_{\max}$ is the maximum velocity, and $T_\sigma$ is the capillary timescale $\sqrt{\rho D^3/\sigma}$. Mesh convergence of the capillary number is examined for simulations with Laplace numbers of $\La=1.2\times 10^2, 1.2\times 10^4$, and $1.2\times 10^6$. 
The spurious velocities are measured across $N_s=50$ random initial droplet positions, and the maximum capillary numbers are calculated as
\begin{align}
    \max\left(\Ca_{\rms}\right)&=\max_{i\in\{1,...,N_s\}}\Ca_{\rms,i} \\
    \max\left(\Ca_{\max}\right)&=\max_{i\in\{1,...,N_s\}}\Ca_{\max,i}.
\end{align}

\begin{figure}[p]
    \centering
    \includegraphics[width=1.0\columnwidth]{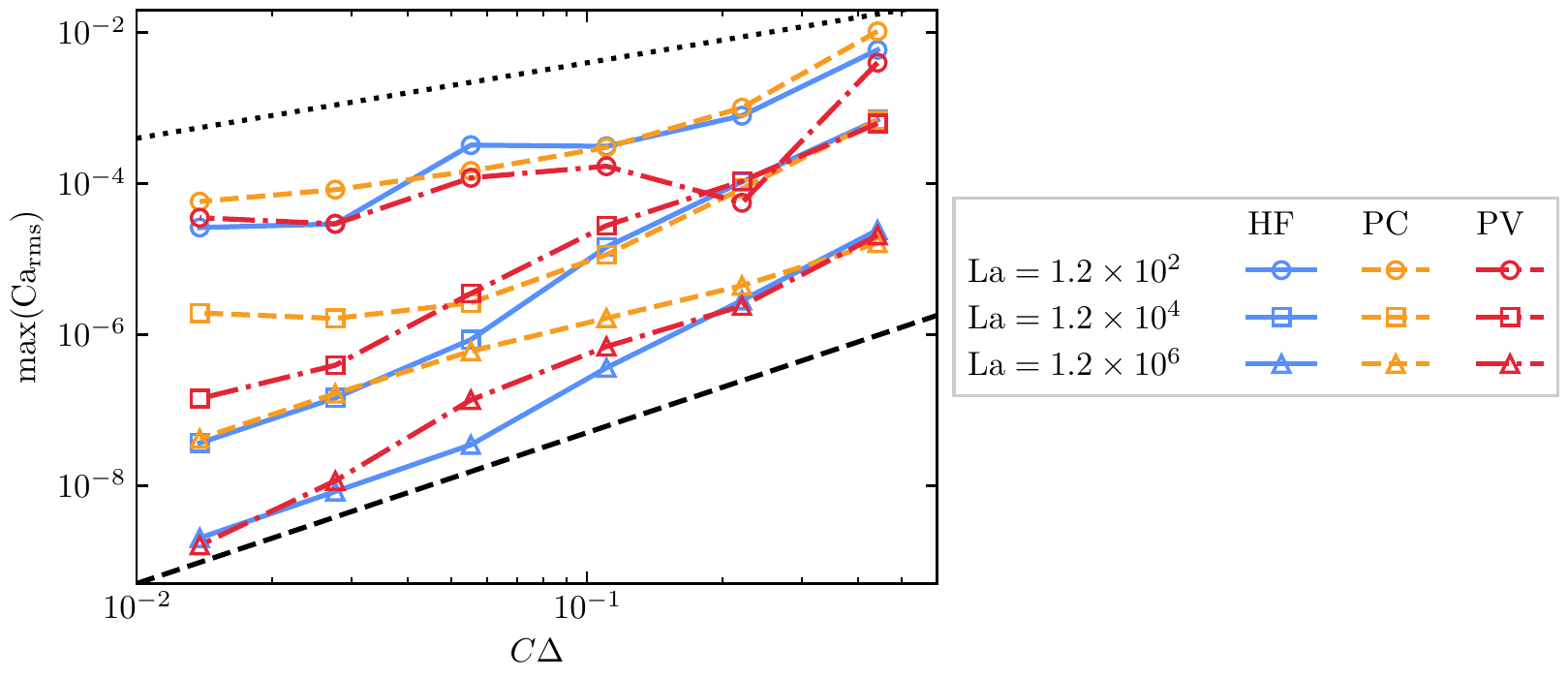}
    \caption{Mesh convergence of capillary number $\max\left(\Ca_{\rms}\right)$ for the 2D stationary droplet test case with curvatures calculated with the height function (HF), PLIC-centroidal (PC), and PLIC-volumetric (PV) methods. The PV method is used for interfacial cells where the HF method fails. Simulations are performed for Laplace numbers $\La=1.2\times 10^2, 1.2\times 10^4$, and $1.2\times 10^6$. The capillary number for each curvature is the maximum of 50 simulations with randomly chosen droplet center. The dotted and dashed lines represent first- and second-order convergence, respectively.}
    \label{fig:stationary-droplet-rms}
\end{figure}

\begin{figure}[p]
    \centering
    \includegraphics[width=1.0\columnwidth]{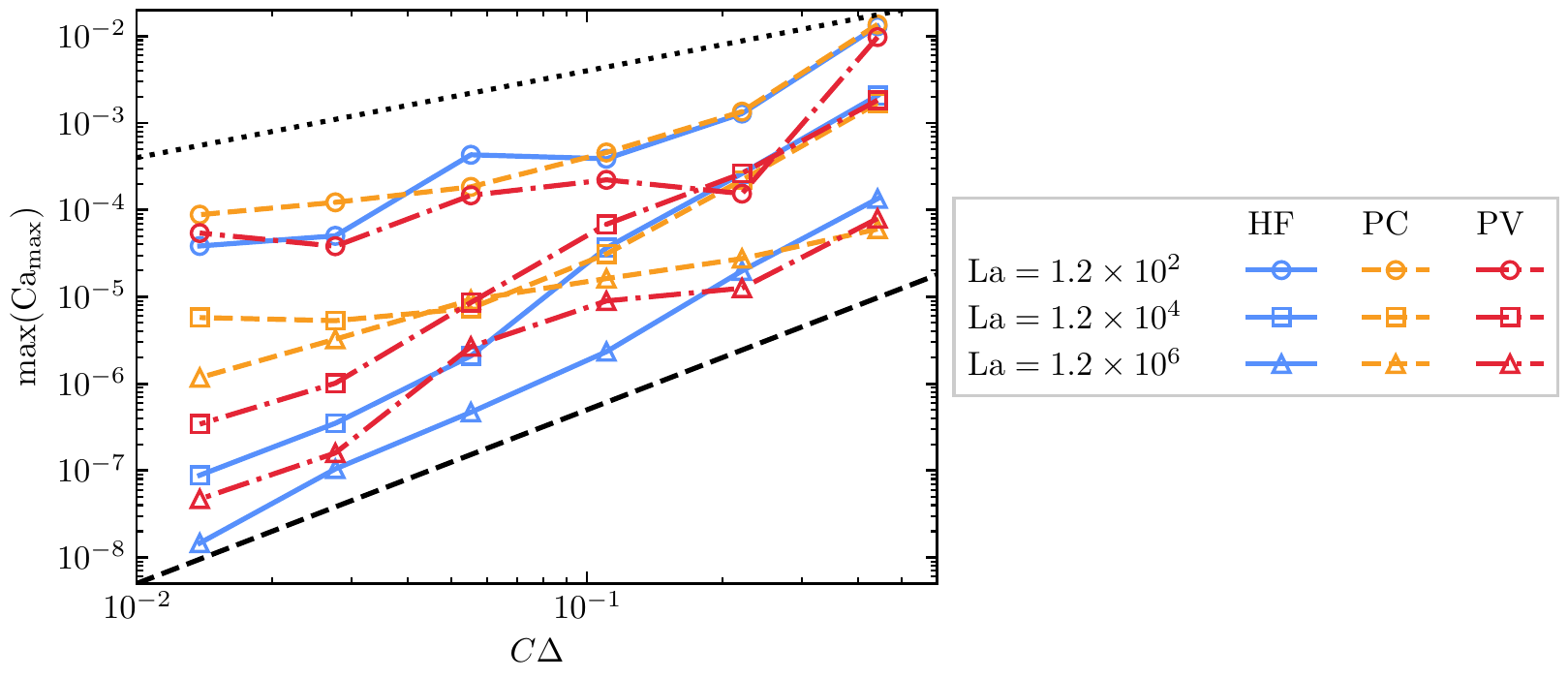}
    \caption{Mesh convergence of capillary number $\max\left(\Ca_{\max}\right)$ for the 2D stationary droplet test case with curvatures calculated with the height function (HF), PLIC-centroidal (PC), and PLIC-volumetric (PV) methods. The PV method is used for interfacial cells where the HF method fails. Simulations are performed for Laplace numbers $\La=1.2\times 10^2, 1.2\times 10^4$, and $1.2\times 10^6$. The capillary number for each curvature is the maximum of 50 simulations with randomly chosen droplet center. The dotted and dashed lines represent first- and second-order convergence, respectively.}
    \label{fig:stationary-droplet-max}
\end{figure}

The mesh convergences of $\max\left(\Ca_{\rms}\right)$ and $\max\left(\Ca_{\max}\right)$ for the HF, PC, and PV methods are respectively displayed in Figures \ref{fig:stationary-droplet-rms} and \ref{fig:stationary-droplet-max} for $D/\Delta$ between 3.2 and 102.4, corresponding to $C\Delta$ of 0.44 and 0.014, respectively. The HF method produces the lowest spurious velocities, while the PV method produces lower spurious velocities than the PC method across all tested Laplace numbers. Note that for all tested $C\Delta$, the HF method fails to estimate a curvature for some of the interfacial cells in the domain. For droplets with $C\Delta=0.44$, the HF method always fails to estimate a curvature regardless of the center location of the droplet, while for droplets with $C\Delta=0.014$, 28\% of the randomly centered droplets required use of the PV method as a backup, with the maximum HF failure rate being 0.9\%. The capillary number converges with second-order accuracy for $\La=1.2\times 10^4$ and $1.2\times 10^6$ using the HF and PV methods. The convergence rate using the PC method is between first- and second-order for $\La=1.2\times 10^6$, and second-order for $\La=1.2\times 10^4$ except with the finest meshes where the errors fail to converge. With $\La=1.2\times 10^2$, the convergence is approximately first-order for all methods, with all methods losing convergence with the finest meshes. The decaying convergence rate of the capillary number with decreasing $C\Delta$ in the $\La=1.2\times 10^2$ case is consistent with the convergence behavior of the curvature error in the random paraboloids case when the volume fractions are randomly perturbed. Likewise, the convergence behavior of the capillary number in the $\La=1.2\times 10^4$ and $\La=1.2\times 10^6$ cases resembles that of the curvature error in the random paraboloids case when the volume fractions are not perturbed, as the HF and PV methods generate both lower spurious velocities in the stationary droplet case and lower curvature errors in the random paraboloids case than the PC method. The stationary droplet case produces results similar to those of the paraboloids with minimally perturbed volume fractions because the droplets do not undergo significant transport and therefore do not incur large errors in the volume fractions. In the following section, the translating droplet case evaluates the performance of the curvature calculation methods in the presence of large volume fraction errors from interfacial transport.

\subsection{3D translating droplet}
This case examines the interaction of the curvature and surface tension calculation with the flow solver when the interface is advected over a length $l \gg \Delta$. A spherical droplet of diameter $D=1\times 10^{-4}$ is placed in the center of a three-dimensional cubic domain of length $L=2.5D$ with periodic boundary conditions and a uniform Cartesian grid of mesh size $\Delta$. The droplet is initialized in a uniform velocity field $\mathbi{u}_0=[u_0,u_0,u_0]$, where $u_0=0.5/\sqrt{3}$ such that $U=|\mathbi{u}_0|=0.5$. The droplet density, $\rho_l=1000$, and viscosity, $\mu_l=1.137\times 10^{-3}$, are chosen to match those of water at standard conditions. The density and viscosity of the surrounding fluid are equal to those of air: $\rho_g=1.226$ and $\mu_g=1.780\times 10^{-5}$. The surface tension coefficient is $\sigma=0.0728$. The spurious velocities are measured with the capillary numbers $\Ca_{\max}=|\mathbi{u}-\mathbi{u}_0|_{\max}\mu/\sigma$ and $\Ca_{\rms}=|\mathbi{u}-\mathbi{u}_0|_{\rms}\mu/\sigma$ evaluated at nondimensional time $tU/D=7.84$. 

\begin{figure}[p]
    \centering
    \includegraphics[width=0.7\columnwidth]{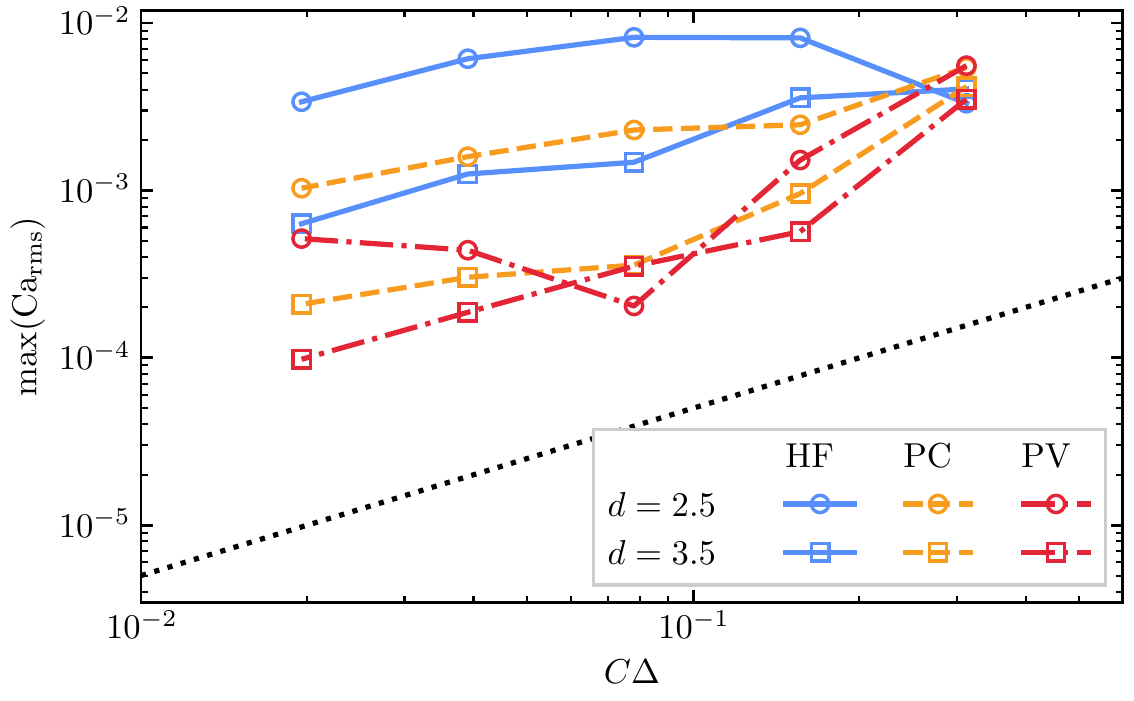}
    \caption{Mesh convergence of capillary number $\Ca_{\rms}$ for the 3D translating droplet test case with curvatures calculated with the height function (HF), PLIC-centroidal (PC), and PLIC-volumetric (PV) methods. The PV method is used for interfacial cells where the HF method fails. The curvature stencil radial weighting width is varied between $d=2.5$ and $d=3.5$, including for the PV method when used as a backup for the HF method. The dotted line represents first-order convergence.}
    \label{fig:translating-droplet-rms}
\end{figure}

\begin{figure}[p] 
    \centering
    \includegraphics[width=0.7\columnwidth]{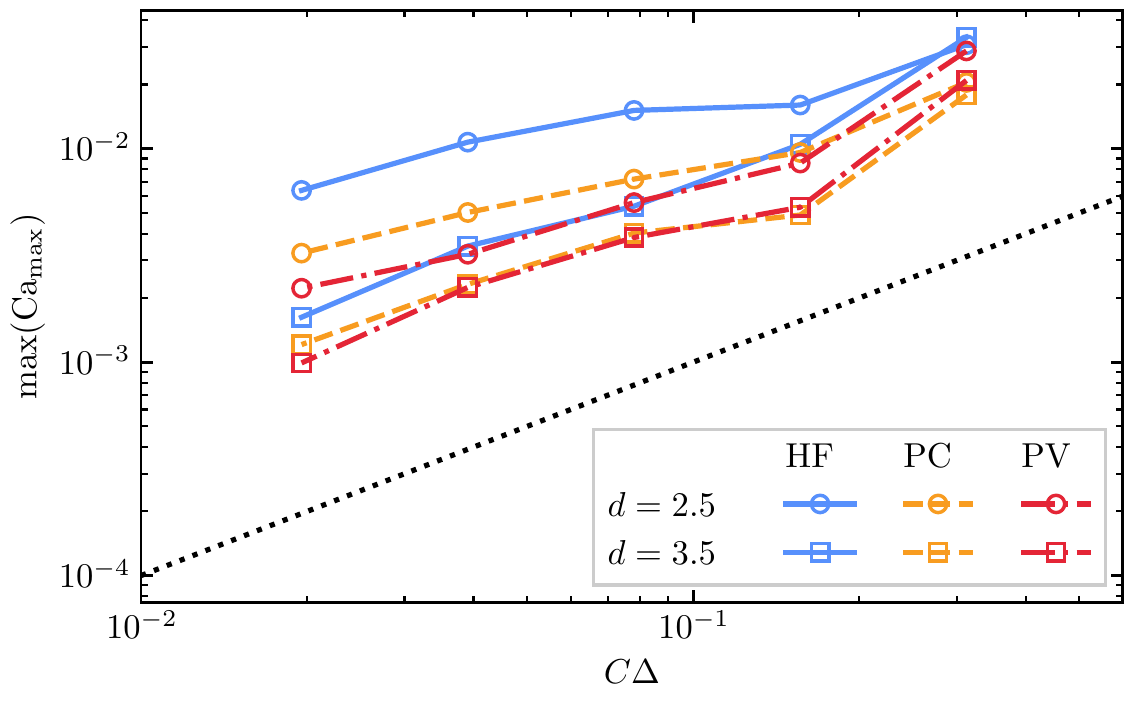}
    \caption{Mesh convergence of capillary number $\Ca_{\max}$ for the 3D translating droplet test case with curvatures calculated with the height function (HF), PLIC-centroidal (PC), and PLIC-volumetric (PV) methods. The PV method is used for interfacial cells where the HF method fails. The curvature stencil radial weighting width is varied between $d=2.5$ and $d=3.5$, including for the PV method when used as a backup for the HF method. The dotted line represents first-order convergence.}
    \label{fig:translating-droplet-max}
\end{figure}

Figures \ref{fig:translating-droplet-rms} and \ref{fig:translating-droplet-max} show the mesh convergence of $\Ca_{\rms}$ and $\Ca_{\max}$, respectively, for the HF, PC, and PV methods for $D/\Delta$ between 6.4 and 102.4, where for the HF method, the PV method is used as a backup, and both $d=2.5$ and $d=3.5$ radial weighting widths are tested. For a spherical droplet, $D/\Delta$ values of 6.4 and 102.4 correspond to $C\Delta$ values of 0.3125 and 0.0195, respectively. With the use of the narrower $d=2.5$ weighting, all methods show below first-order convergence with capillary numbers being lower for the PV method with finer meshes and the HF method producing much higher $\Ca_{\rms}$ than the other methods. When the radial weighting width is increased to $d=3.5$, first-order convergence is recovered for all methods with capillary numbers being lower for the PV method with finer meshes and the HF method producing much higher $\Ca_{\rms}$ than the other methods. All of the methods produce similar spurious velocities with the $D/\Delta=6.4$ droplet. 
Note that if the mesh resolution were to be further increased, the convergence would probably decay to below first-order, as predicted by Eq.\ \eqref{convergence_order}. The decay in mesh convergence occurs at higher values of $C\Delta$ than for the stationary droplet because of the increased accrual of volume fraction errors due to the transport of the volume fractions across multiple cell lengths. Just as in the random paraboloids test, the HF method produces higher errors than the PC and PV methods when the volume fractions incur significant errors. The translating droplet case also highlights the dependence of the spurious velocities on the HF backup method, as the difference in capillary numbers when using the combined HF and PV method with a $d=2.5$ weight versus using a $d=3.5$ weight mirrors that between the standalone PV methods with a $d=2.5$ weight and $d=3.5$ weight. For a given radial weighting width $d$, the PV method produces lower spurious currents than those from the PC and HF methods.

Mesh convergence does not improve monotonically with increasing $d$, however, as when the weighting is widened to $d=4.5$, the capillary number fails to converge with mesh refinement. This is because, as noted in \cite{Owkes2018}, the radial weighting of the curvature fit effectively smooths the interface to decrease the sensitivity of the fit to erroneous fluctuations in the volume fraction from the transport and reconstruction steps. However, too wide of a weighting width can flatten small interfacial perturbations that result from local spatial velocity fluctuations that exist at the scale of the mesh size, leading to an underestimation of the local surface tension force. The choice of a radial weighting width is therefore a balance between the locality of the curvature estimation and the robustness of the estimation to volume fraction errors. The results presented in this work, along with those from \cite{Owkes2018}, show that the curvature fitting should give the largest weight to data located within one to two cell lengths of the curvature estimation location. 

\section{Conclusions}\label{conclusions}
This work compares the performance of four curvature estimation methods for interfaces represented implicitly by a discrete volume fraction field: the height function method \cite{Hirt1981}, a paraboloid fit to interface reconstruction centroids, the reconstruction-based volumetric fitting method of Jibben et al. \cite{Jibben2019}, and a novel direct volumetric fitting method. The first three methods are evaluated with both static and dynamic interfaces in two and three dimensions.

The test results demonstrate that the method of Jibben et al. best balances low curvature errors with low computational cost in realistic cases where the volume fractions in interfacial cells incur errors, such as those from the transport and reconstruction steps in interfacial flow simulations using geometric volume-of-fluid methods. In the dynamic interface tests, the method of Jibben et al. results in lower spurious velocities than the height function method over a large range of mesh resolutions. While the proposed fully volumetric fitting method produces curvature errors that converge with mesh refinement with second-order accuracy when exact volume fractions are used, it loses accuracy when first-order volume fractions are used such that the errors are comparable to those from the method of Jibben et al. The fully volumetric fitting method, however, could be the basis of a higher-order interface reconstruction method that would produce the second-order volume fractions necessary for mesh convergence of curvature error over all mesh resolutions. The results of this work also highlight the importance of including realistic cases incorporating interface transport in curvature evaluation tests, as the relative performance of the methods in these cases greatly differs from that in cases where the interface does not undergo significant transport.

One limitation of the curvature estimation tests presented in this work is that they only evaluate performance on discrete volume fractions in a Cartesian grid. Further testing should be performed on a variety of unstructured meshes to ascertain the applicability of the presented results to non-Cartesian meshes.

\section{Acknowledgements}
This work was sponsored by the Office of Naval Research (ONR) as part of the Multidisciplinary University Research Initiatives (MURI) Program, under grant number N00014-16-1-2617.  The views and conclusions contained herein are those of the authors only and should not be interpreted as representing those of ONR, the U.S. Navy, or the U.S. Government. 

F. Evrard is funded by the European Union’s Horizon 2020 research and innovation program under the Marie Skłodowska-Curie Grant Agreement No. 101026017.
\bibliography{allreferences}
\end{document}

%% file: commonm.tex
\usepackage{amsfonts}
\usepackage{amsmath}
\usepackage{amssymb}
\usepackage{color}
\usepackage{tikz}
\usepackage{pgfplots}
\usepackage{listings}
\usepackage{courier}
\usepackage{siunitx}
\usepackage{array}
\usepackage{svg}

\def\mathbi#1{\textbf{\em #1}}

\newcommand{\calU}{\mathcal{U}}
\newcommand{\calN}{\mathcal{N}}

\newcommand{\La}{\operatorname{La}}
\newcommand{\Ca}{\operatorname{Ca}}
\newcommand{\rms}{\operatorname{rms}}

\newcommand{\vertiii}[1]{{\left\vert\kern-0.25ex\left\vert\kern-0.25ex\left\vert #1
    \right\vert\kern-0.25ex\right\vert\kern-0.25ex\right\vert}}

\newcommand{\order}[1]{\mathcal{O}( #1 )}

\newcommand{\ltwo}{L_2}
\newcommand{\linf}{L_{\infty}}